# Lifetime measurement of the low lying yrast states in $^{189}$Pt

HE Chuang-Ye(贺创业) WU Xiao-Guang(吴晓光) WANG Jin-Long (汪金龙) WU Yi-Heng(吴义恒)　ZHENG Yun(郑云) LI Guang-Sheng(李广生) LI Cong-Bo(李聪博) HU Shi-Peng(胡世鹏) LI Hong-Wei(李红伟) LIU Jia-Jian(刘嘉健) LUO Peng-Wei(罗朋威) YAO Shun-He(姚顺河)

1 China Institute of Atomic Energy, Beijing 102413, China

## abstract

Lifetimes of the positive-parity yrast band in $^{189}$Pt were measured using the recoil distance Doppler-shift method. A HPGe detector array consisting of 13 detectors was used in conjunction with the plunger device in CIAE. Excited states of $^{189}$Pt were populated by the $^{176}$Yb ($^{18}$O, 5n) $^{189}$Pt fusion-evaporation reaction at a beam energy 87 MeV. The lifetimes of two levels belonging to the yrast band are measured. The results show that the $17/2^{+}$ state in the yrast band has large $Q_t$ value, but it deceases quickly with spin increasing. It may be contributed from the shape driving effect of the quasi-neutron from the $i_{13/2}$ oribital.

## 1. Introduction

Nuclear shapes of the Pt-Au-Hg transitional region are very unstable with respect to γ-deformation. This region is known to exhibit triaxiality [1] and shape coexistence [1-5]. The occupation of different subshells of a specific high-j orbitals near the Fermi surface can induce shape changes for Such "γ-soft" nuclei. For the Pt isotope chain, nuclei have a shape transition from triaxial-prolate to triaxial-oblate around A = 187 with neutron number increasing. The ground states of odd Pt nuclei are prolately deformed for A ≤ 187, while for the nuclei of A ≥ 189 are oblately deformed. It is very interesting to study nuclear shape evolution in those nuclei at this point. The $^{189}$Pt [6] nuclei have been studied via $^{176}$Yb($^{18}$O, 5n) $^{189}$Pt reaction at beam energies of 88 and 95 MeV. Some features of their level spectra suggest a deviation from axial symmetry for the positive-parity yrast band. It has also been pointed out that the complex low-lying positive-parity level structures can be described [6] assuming a coupling of the high-j ν $i_{13/2}$ hole to a rotating triaxial core. $^{189}$Pt nuclei are therefore a good example to study the high j shape driving effect of $i_{13/2}$ orbital.

In this paper, we report on the first measurement of nuclear lifetimes in the picosecond range using the Recoil Distance Doppler Shift method (RDDS) developed for heavy ion fusion-evaporation reactions in combination with HPGe detector array in China Institute of Atomic Energy (CIAE), Beijing. The lifetimes of the low spin states in the yrast band of $^{189}$Pt have been determined.

## 2. Experimental Details

High spin states in $^{189}$Pt were populated through the fusion-evaporation reaction $^{176}$Yb($^{18}$O, 5n)$^{189}$Pt at a beam energy of 87 MeV. The $^{18}$O beam was delivered by the HI-13 tandem accelerator of CIAE, Beijing. The target consisted of 99% enriched $^{176}$Yb isotope material evaporated to a thickness of 0.7 mg/cm$^2$ onto a 2 mg/cm$^2$ thick Au foil. The recoiling nuclei left the target with a mean velocity v of 0.94% of the velocity of light c and were stopped in a 3 mg/cm$^2$ Au foil mounted together with the target in the plunger apparatus. Data were collected for 7 target-to-stopper distances ranging from 1 to 6000 μm. The accuracy of the measurement of the relative target-to-stopper distances was better than 0.1 μm in the range from electrical contact to 30 μm and better than 0.5 μm in the range 30–200 μm.

In this experiment, a array consisted of 11 Compton suppressed HPGe detectors and two planer-type HPGe detectors was used.   Four of these detectors were placed at 90°, two at 140° and 150°, and one each at 42°, 45°, 50°, 122°, and 145° respect to the beam direction. However, the detectors placed around 90° with respect to the beam direction axis could not be used to measure lifetimes since the Doppler shift of a γ ray emitted in flight is close to zero for these detectors. Details of nuclear level lifetime measurement by using Plunger in CIAE in combination with the RDDS method has been previously reported in Ref. [7].

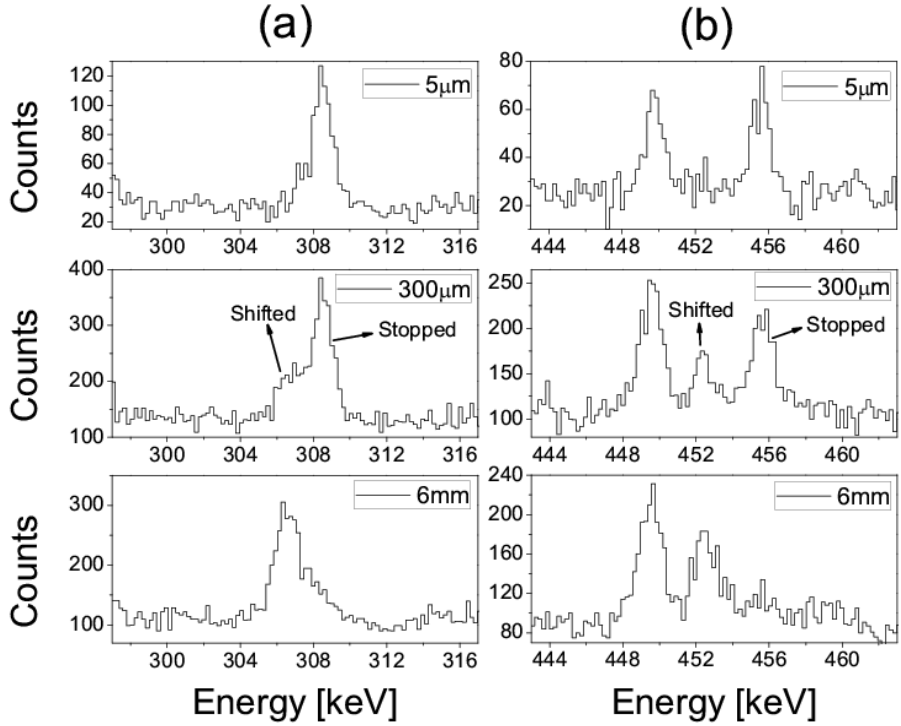

Fig.1: Sample spectra from the "backward-angle" detectors for $^{189}$Pt at the indicated recoil distances. The (a) and (b) spectra show the $17/2^+\rightarrow13/2^+$ 308-keV and the $21/2^+\rightarrow17/2^+$ 456-keV transitions at different target-stopper distances. The relative increase in the intensities of the "shifted" peaks with increasing distance is clearly visible.

## 3. Results and Discussion

In this work, lifetime measurements have been made for the two lower-spin states in the yrast band of $^{189}$Pt. Lifetimes were determined from observing the ratio of the fully Doppler-shifted and stopped peaks from data collected by the detectors at a backward angle of 150° to the beam line. Figure 1 shows, for different target-stopper distances, Doppler-shifted γ-ray spectra associated with $^{189}$Pt for selected energy regions including their respective $17/2^+ \rightarrow 13/2^+$ 308-keV and $21/2^+ \rightarrow 17/2^+$ 456-keV transitions. The γ-ray energy spectra measured at backward angles for distances of 5, 300, and 6000 μm are presented. One can clearly see the shifted and unshifted components of the 308, and 456 keV transitions in $^{189}$Pt. The peak areas of both components were determined as a function of target-to-stopper distance. The ratio of the shifted (stopped) peak area to the total area (sum of shifted and stopped) was determined for each distance. The data analysis method has been described in detail in Ref. [7,8]. The lifetime of $17/2^+$ and $21/2^+$ states were finally obtained, as list in Table I.

Table.1 Measured lifetimes and the corresponding $Q_t$ values for the yrast band in the $^{189}$Pt.

| Iπ | $E_γ$(MeV) | τ(ps) | $Q_t$(eb) |
|---|---|---|---|
| $17/2^+$ | 308 | 305(8) | 11.05(82) |
| $21/2^+$ | 456 | 253(7) | 2.49(17) |

The transition quadrupole moment ($Q_t$, in eb) is related with the level lifetimes (τ, in ps) through the expression

$$\tau = \frac{16\pi}{12.24 \times 5 < IK20|I-2K >^2} \frac{f_γ(E2, I \rightarrow I-2)}{E_γ^5(I \rightarrow I-2)Q_t^2}$$

(1)

where the γ-ray energy $E_γ$ in MeV. The Clebsch-Gordan coefficient can be written as

$$< IK20|I-2K >^2 = \frac{3}{8} \frac{I(I-1)}{I^2 - 1/4} \frac{I^2 - K^2}{I^2} \frac{(I-1)^2 - K^2}{(I-1)^2}$$

(2)

where $f_γ$(E2, I→I-2) is the branching ratio of the E2 transition to all γ transitions from the level. According to the relationship between the quadrupole deformation parameter β and the transition quadrupole moment $Q_t$,

$$Q_t = \frac{3}{\sqrt{5\pi}} Zr_0^2 A^{\frac{2}{3}} (\beta + \sqrt{\frac{5}{64\pi}} \beta^2)$$

(3)

In the odd-A neutron-deficient Pt isotopes, a yrast band built on $i_{13/2}$ configuration has been widely observed. The previous results show that they generally have a isomer at the band head position. With neutron number increasing, isomers at the band-head were confirmed in the experiment for the isotopes with neutrons number larger than 109 [9-15]. The nature of these bands is of particularly interesting. In the present experiment, the yrast band for $^{189}$Pt has been studied. The results show that the $17/2^+$ state in $^{189}$Pt has a large $Q_t$ value. With the neutron number 111, the neutron Fermi level in $^{189}$Pt is considered to lie at the high $\Omega$ subshell of $i_{13/2}$ orbital. In reference [6], the $13/2^+$ bandhead is explained by the occupation of $i_{13/2}$ 11/2[615] orbital. Valence particles from this orbital have large oblate shape driving effect. On the other hand, the yrast band in $^{189}$Pt could be described as a $i_{13/2}$ neutron-hole weakly coupled with the oblate deformed nucleus $^{190}$Pt core resulting from the similar level spacings with the ground band in $^{190}$Pt. This means that the deformation of the $17/2^+$ is higher than the deformation of the $2^+$ state since it is associated with the $i_{13/2}$ structure which drives the nucleus into a more oblate deformed shape. From Eq. [3], we can get the $\beta$ deformation for $17/2^+$ state is -0.28. It is much larger than the value of -0.156 for the ground state of $^{190}$Pt [16]. It is also larger than -0.164 [16] for the ground state of $^{189}$Pt. This is maybe one of the reason why the isomer come into being at $13/2^+$ state.

As referred before, the yrast band in $^{189}$Pt is suggest as a oblate band with a considerable $\beta$ deformation at the band head position. However, the $\beta$ deformation decreases quickly with spin increasing. The rotational hypothesis can be tested by comparing the moments of inertia $J^{(1)}$ [17,18], which is often used as a measure of deformation of collectivity in nuclear rotational bands, inferred from the experimental energies and spins with those of the yrast bands of nearby nuclei. Classically, in terms of the spin I and rotational frequency, $J^{(1)}$ can be given by $J^{(1)}=I/\omega$. The $\omega$ is related to this derivative: $\omega=dE/dI$, where E is the excitation energy corresponding to each state. The derivative can be estimated by the ratio of finite differences of energies and spins between two adjacent levels, $\frac{dE}{dI} = [E(I) - E(I-2)]/[\sqrt{I(I+1)} - \sqrt{(I-1)(I-2)}]$ Fig. 2 shows that the $J^{(1)}$ values decrease considerably with angular momentum increasing, and the decreasing extent diminishes at higher spin states. This behavior is generally consistent with the results from lifetime measurement. It is not surprising that the deformation will decreasing with spin. The particle plus rotor model calculation [6] shows that the valence neutron occupation will change form $i_{13/2}$ 11/2[615] to $i_{13/2}$ 1/2[660] orbital at higher spin states. Valence particle from this orbital has large prolate shape driving effect. So the nuclei tend to be driven into a spherical shape. Then the β deformation will be decreased.

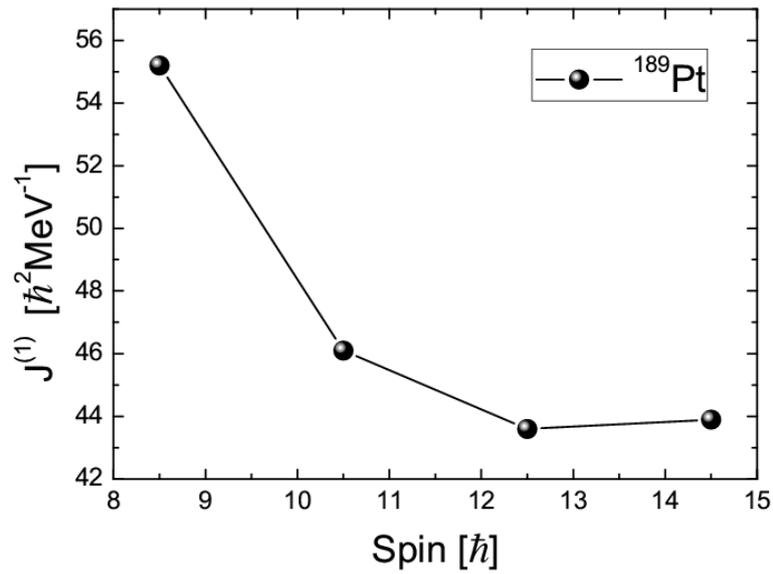

Fig2: Kinematic moments of inertia $J^{(1)}$ as a function of spin for low spin states of the positive-parity yrast band in $^{189}$Pt.

## 4. Summary

In summary, the lifetimes of two excited states in $^{189}$Pt have been determined using the Recoil Distance Doppler Shift method with the HPGe detector array in CIAE. The results show that the first excited state in the yrast band has large $Q_t$ value, but it deceases quickly with spin increasing. Compared with the β deformation of the ground state, $17/2^+$ state is much more deformed. The behavior of the decreasing $J^{(1)}$ values generally agree with the results of lifetime measurement.

## 5. Acknowledgement

We would like to thank the staff of the HI-13 tandem accelerator in the China Institute of Atomic Energy for steady Providing the $^{18}$O beam. The authors are greatly indebted to DR. Q. W. Fan and Y. H. Du for preparing the target. This work has been partially supported by the NSFC (Grant Nos. 11175259, 10927507, 11075214).